\documentclass{IEEEtran}
\usepackage[nocompress]{cite}
\usepackage{amsmath,amssymb,amsfonts}
\usepackage[caption=false,font=footnotesize]{subfig}
\usepackage{algorithmic}
\usepackage{graphicx}
\usepackage{textcomp}
\usepackage{orcidlink}
\usepackage{multirow}
\usepackage{xparse} 
\usepackage{acro} 
\usepackage{stfloats} 
\usepackage{capt-of}

\newcommand{\junit}{\mathrm{j}}

\newcommand{\op}[1]{\mathcal{#1}}
\DeclareAcronym{gsie}{
    short = GSIE,
    long  = generalized source integral equation
}

\DeclareAcronym{sie}{
    short = SIE,
    long  = surface integral equation
}

\DeclareAcronym{mom}{
    short = MoM,
    long  = method of moments
}

\DeclareAcronym{pmchwt}{
    short = PMCHWT,
    long  = Poggio-Miller-Chang-Harrington-Wu-Tsai
}

\DeclareAcronym{lr}{
    short = LR,
    long  = low-rank
}

\DeclareAcronym{mgk}{
    short = MGK,
    long  = modified Green's kernel
}

\DeclareAcronym{mgf}{
    short = MGF,
    long  = modified Green's function
}

\DeclareAcronym{tm}{
    short = TM,
    long  = transverse magnetic
}

\DeclareAcronym{te}{
    short = TE,
    long  = transverse electric
}

\DeclareAcronym{dof}{
    short = DoF,
    short-plural-form = DoFs,
    long = degree of freedom,
    long-plural-form = degrees of freedom
}

\def\BibTeX{{\rm B\kern-.05em{\sc i\kern-.025em b}\kern-.08em
    T\kern-.1667em\lower.7ex\hbox{E}\kern-.125emX}}

\begin{document}
\bstctlcite{IEEEexample:BSTcontrol}
\title{A Generalized Source Integral Equation for Homogeneous Penetrable Scatterers}
\author{Boris Diner\,\orcidlink{0009-0007-1732-852X}, \IEEEmembership{Graduate Student Member, IEEE}, and Yaniv Brick\,\orcidlink{0000-0003-1026-8273}, \IEEEmembership{Senior Member, IEEE}
\thanks{{
This work was supported in part by Israel Science Foundation (ISF)
under Grants 442/22 and 411/26.} (\textit{Corresponding author: Yaniv Brick})}
\thanks{
The authors are with the School of Electrical and Computer Engineering, Ben-Gurion University of the Negev, Beersheba 8410501, Israel (e-mail: ybrick@bgu.ac.il).}}

\maketitle

\begin{abstract}
The \ac{gsie} approach is extended to enhance the compressibility of moment matrices for essentially-convex homogeneous penetrable scatterers. In the proposed indirect \ac{sie} formulation, the scattered field representations use a single distribution of generalized sources that radiate only weakly into the scatterer. A separate distribution of conventional sources is used for representing the interior fields. The wave interactions between surface subdomains via the generalized kernel are of a reduced effective dimensionality. The corresponding exterior integral operator moment matrix blocks exhibit enhanced rank deficiency.  For lossy scatterers, this secures the \ac{lr} compressibility of hierarchical matrix structures for the full system. For scatterers in a much slower background medium, significant savings with very delayed asymptotic scaling of the costs can be achieved. The \ac{gsie} is formulated for the \ac{tm} scattering problem. It is validated for representative examples and its superior compressibility is demonstrated.
\end{abstract}


\section{Introduction}
\label{sec:introduction}
\IEEEPARstart{S}{cattering} problems for homogeneous penetrable objects (e.g.,~\cite{ groth2015boundary, sukharevsky2017scattering, dehmollaian2019wave}) can be formulated as \acp{sie}. The conventional direct \acp{sie} use equivalence principle descriptions of the fields in the exterior and interior problem regions. The unknown electric and magnetic fields are expressed via integrals over suitable traces and their continuity across the surface is imposed~\cite{poggio1973integral, muller1969foundations, yla2008analysis, mavrikakis2025surface, le2026multitracemuller}. Upon their discretization, e.g., by the \ac{mom}~\cite{harrington1968field}, the \ac{sie} systems are translated to systems of linear equations characterized by large and dense system-matrix blocks, which correspond to combinations of integral operators of different kernels.
Their storage and solution for electrically-large objects are prohibitively expensive, thus calling for efficient matrix-free (e.g.,~\cite{song1995multilevel, bleszynski1996aim, wei2014more, phillips1997precorrected, brick2010multilevel}) or algebraically-compressed (see~\cite{canning2003universal, zhao2005adaptive, shaeffer2008direct, brick2011fast, li2014nested, michielssen1996multilevel, guo2018butterfly, heldring2026butterfly}) representations that can be used in fast iterative and direct solvers. 

For these formulations, kernel-independent algebraic compression is of particular appeal. The discretized \ac{sie} operators are often partitioned hierarchically and their potentially-compressible blocks are expressed economically (e.g.,~\cite{hackbusch1999sparse, borm2003introduction, bebendorf2008hierarchical, martinsson2025fastdirectsolvers}). If the blocks are rank-deficient, their \ac{lr} approximation leads to significant savings and, in some cases, reduction of the asymptotic scaling of costs with the electrical length  $kR$. For example, for reduced dimensionality (e.g., elongated~\cite{michielssen1996scattering, martinsson2007fast} and quasi-planar~\cite{ corona2015direct,gaede2025fast}) geometries, the ranks scale sublinearly with the number of problem unknowns $N$. For general surfaces, the broadside interactions between surface subdomains are of numbers of \acp{dof} that scale asymptotically with $kR$ such that the blocks ranks are of $\mathcal O(N)$  \cite{bucci1989degrees, brick2016fast, brick2018rapid, brick2026interpreting}\footnotemark. This prevents the reduction of the asymptotic costs via \ac{lr} compression.
 \footnotetext{This is also the case for the discretized operators in single-source
formulations that use conventional \ac{sie} kernels, such as those in ~\cite{menshov2012new,lori2017novel,patel2017novel}.}

To overcome this limitation, \acp{gsie} were proposed~\cite{boag2012generalized, brick2014fast,sharshevsky2020direct, zvulun2023generalized, dahan2024fast, dahan2026vector, kalhofer2025fast, kalhofer2026dyadic}. In these \acp{sie}, introduced for impenetrable essentially-convex geometries, the scattered fields are represented by integrals on generalized source distributions that radiate according to \ac{mgf} kernels. These kernels are composed of the conventional free-space Green's function contributions and auxiliary ones designed to emulate sources located within the scatterer. The latter suppress the broadside and accentuate the end-fire components of the interactions between subdomains. The resulting interactions are of reduced effective dimensionality. The corresponding block ranks and the associated solver costs scale significantly slower with $kR$~\cite{brick2014fast, dahan2024fast, dahan2026vector, kalhofer2025fast}. These equations also exhibit promising spectral properties and respond well to preconditioning~\cite{dahan2026wellconditioned}.  However, the application of the approach to penetrable objects is non-trivial.

This work extends the \ac{gsie} approach to homogeneous penetrable scatterers. To this end, the problem is formulated indirectly, using generalized sources for representing the (external) scattered fields and conventional sources for the internal fields. For each region, a single source distribution is used. An \ac{sie} system results from imposing the continuity of tangential fields. The system blocks associated with the external region exhibit enhanced compressibility. If the remaining system blocks, which are associated with the internal region, correspond to either lossy or much faster material, they are inherently highly-compressible and may only (for latter) govern the costs for very large $kR$ values. This is in contrast with conventional formulations, where the internal and external blocks are combined into structures of fast scaling ranks dictated by the (conventional) kernel in the lossless slow medium. This concept, outlined in~\cite{diner2026compressible}, is validated here for \ac{tm} problems and multipole-based \ac{gsie} kernels \cite{kalhofer2025fast}, which enable studying it for various types of geometries. 

The remainder of this letter is organized as follows: Section~II presents the proposed \ac{gsie} formulation. Section~III presents numerical results that validate it and demonstrate its enhanced compressibility. Section~IV concludes the work.
\begin{figure*}[!b]
    \includegraphics[]{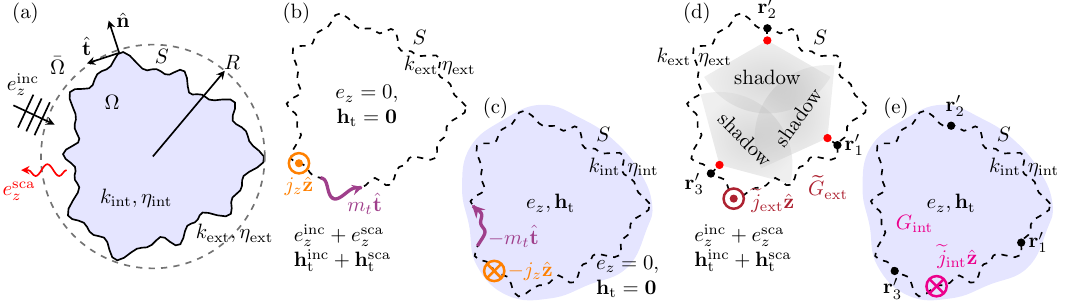}
    \caption{(a) Scattering problem. (b), (c) Equivalent problems for $\bar \Omega $ and $\Omega$, respectively. (d), (e)  Generalized source representation in $\bar \Omega $ and $\Omega$, respectively.}
     \label{fig:penetrable_scattering}
\end{figure*}
\section{Generalized \ac{sie} Formulations}
\label{sec:formulation}
Consider the scattering of \ac{tm} incident fields ${\mathbf e}^{\mathrm{inc}}=e_z^{\mathrm{inc}}{\hat {\mathbf z}}$ and ${\mathbf h}^{\mathrm{inc}}={\mathbf h}_{\mathrm t}^{\mathrm{inc}}$ by a $z$-invariant homogeneous scatterer defined by an essentially- (i.e., a small deviation from) convex surface $S$, with normal and tangent vectors $\hat {\mathbf n}$ and $\hat {\mathbf t}$, bounding a volume $\Omega$ of typical dimension $R$, as illustrated in Fig.~\ref{fig:penetrable_scattering}(a). An $e^{\junit\omega t}$ time dependence is assumed and suppressed. The exterior and interior volumes $\bar \Omega = \mathbb{R} \backslash \Omega$ and $\Omega$ are characterized by wavenumber and impedance pairs $(k_i,\eta_i)$, $i\in\{\mathrm{ext},\mathrm{int}\}$. For ease of notation, let $ k := k_{\mathrm {ext}}$ and $\lambda:= \lambda_{\mathrm {ext}} = 2\pi/k$. 

The scattering problem is conventionally formulated (e.g., in \cite{poggio1973integral,muller1969foundations}) using the equivalence principle, for the scattered and total fields in $\bar \Omega$ and $\Omega$, respectively [see Figs.~\ref{fig:penetrable_scattering}(b) and~\ref{fig:penetrable_scattering}(c)]. From the continuity of the tangential field components $e_z$ and $h_t=\mathbf h \cdot \hat {\mathbf t}$ across $S$ results in a system of \acp{sie}
\begin{equation}
\resizebox{0.9\columnwidth}{!}{$
\begin{bmatrix}
\op{S}_{\mathrm{ext}}+\op{S}_{\mathrm{int}}
&
-\left(\op{D}_{\mathrm{ext}}+\op{D}_{\mathrm{int}}\right)
\\
\op{D}_{\mathrm{ext}}^{*}+\op{D}_{\mathrm{int}}^{*}
&
\op{N}_{\mathrm{ext}}+\op{N}_{\mathrm{int}}
\end{bmatrix}
\begin{bmatrix}
j_z\\
m_t
\end{bmatrix}
=
\begin{bmatrix}
e_z^{\mathrm{inc}}\\
h_t^{\mathrm{inc}}
\end{bmatrix},
$}
\label{eq:pmchwt}
\end{equation}
for the scalar values of the electric and magnetic equivalent surface currents $\mathbf{j} = j_z\mathbf{\hat z}$ and $\mathbf{m} = m_t\mathbf{\hat t} $ on $S$, which are the traces of $\mathbf h$ and $\mathbf e$, respectively. 
The single-, double-, adjoint double- layer, and hypersingular surface integral operators are 
\begin{flalign}
& \op{S}_{i}\{f\}(\mathbf{r})
=
\junit k_i\eta_i
\int_S
G_i(\mathbf r,\mathbf r')f(\mathbf r')\,\mathrm{d}\mathbf r',
&& \mathbf{r}\in S\\
& \op{D}_{i}\{f\}(\mathbf{r})
=
\int_S
\partial_{n'} G_i(\mathbf r,\mathbf r')
f(\mathbf r')\,\mathrm{d}\mathbf r',
&& \mathbf{r}\in S\\
& \op{D}_{i}^{*}\{f\}(\mathbf{r})
=
\int_S
\partial_{n} G_i(\mathbf r,\mathbf r')
f(\mathbf r')\,\mathrm{d}\mathbf r',
&& \mathbf{r}\in S\\
& \op{N}_{i}\{f\}(\mathbf{r})
=
\junit(k_i\eta_i)^{-1}
\int_S
\partial^2_{nn'} G_i(\mathbf r,\mathbf r')
f(\mathbf r')\,\mathrm{d}\mathbf r',
&& \mathbf{r}\in S,
\end{flalign}
respectively. Here, $\partial _{\{n,n'\}} $ denotes the normal derivative at $\{\mathbf{r},\mathbf{r}'\}\in S$ and $G_i(\mathbf{r},\mathbf{r}')= {\mathrm H}_{0}^{(2)}(k_{i}|\mathbf{r} - \mathbf{r}'|)/{4\junit}$ is the Green's function in the homogeneous medium $i$, where ${\mathrm H}_{n}^{(2)}(\cdot)$ is the Hankel function of the second kind and $n$th order. The singular surface integrals are understood in the principal value sense.

In the discretization of~\eqref{eq:pmchwt}, using $2N$ basis and testing functions, the operators are translated to $N \times N$ blocks denoted $\mathbf S_{i}$,  $\mathbf D_{i}$, $\mathbf D_{i}^{*}$, and $\mathbf N_{i}$. 

\begin{figure}[!b]
    \centering

    \captionsetup[subfloat]{labelformat=empty}
    \subfloat[]{%
        \includegraphics[width=0.5\textwidth]{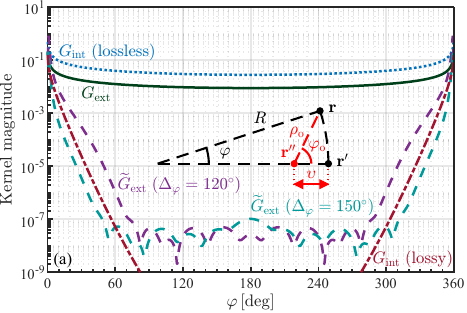}%
        \label{fig:greens_functions}%
    }
    \vspace{-4.5mm}
    \subfloat[]{%
        \includegraphics[width=0.5\textwidth]{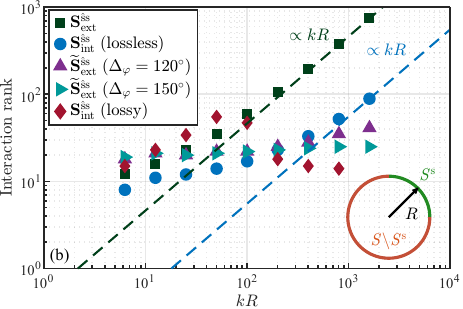}%
        \label{fig:compression_ranks}%
    }

    \caption{(a) Kernel magnitudes on a circular contour for various single-layer-type
    kernels. (b) Corresponding off-diagonal block ranks scaling.}
    \label{fig:greens_rank_combined}
\end{figure}

The exterior-interior block combinations constituting the system blocks in the discretized form of~\eqref{eq:pmchwt} are incompressible: assuming $k_\mathrm{ext}$ describes the slower medium, the ranks of their compressed blocks are governed by the broadside conventional-kernel interactions. For an $(N-N^{\mathrm s}) \times N^{\mathrm s} $ block $\mathbf S_{\mathrm {ext}}^{\mathrm {\hat s s}}$ of $\mathbf S_{\mathrm {ext}}$ that describes an interactions between a large subdomain $S^{\mathrm s}$ of $S$ and the complementary $S \backslash S^{\mathrm s}$, the rank is, asymptotically, of $\mathcal O (kR) =\mathcal O{(N)}$ \cite{brick2026interpreting}. The ranks for all other system blocks scale similarly. 

In the proposed indirect \ac{gsie} formulation, $e_z^{\mathrm {sca}}$ and $h_t^{\mathrm {sca}}$ are integrals over distributions of generalized sources that radiate weakly into $\Omega$ [see  Fig.~\ref{fig:penetrable_scattering}(d)], with kernels related to a \ac{mgf}
\begin{equation}
\widetilde{G}_{\mathrm{ext}}(\mathbf r,\mathbf r')
=
G_{\mathrm{ext}}(\mathbf r,\mathbf r')
+
G_{\mathrm{ext}}^{\mathrm{aux}}(\mathbf r,\mathbf r').
\label{eq:Gout_m}
\end{equation}
The auxiliary contribution $G_{\mathrm{ext}}^{\mathrm{aux}}$ emanates from within $\Omega$ and approximately cancels the radiation into $\Omega$ by the source at $\mathbf r'$. 
For the reduced effective dimensionality of the interactions via the \ac{mgf} to manifest in slower scaling block ranks, the scattered and total fields in $\bar \Omega$ and $\Omega$ are represented by different distributions $\widetilde{j}_{\mathrm{ext}}$ and $\widetilde{j}_{\mathrm{int}}$, respectively. 
The fields inside $\Omega$, which are not well described by integrals with kernels of the nature of~\eqref{eq:Gout_m},  
are expressed using conventional kernels [see Fig.~\ref{fig:penetrable_scattering}(e)]. 
The \ac{sie} system is obtained from continuity of $e_z$ and $h_t$ across $S$.
Defining the generalized single- and adjoint double-layer operators on $S$, 
\begin{flalign}
\widetilde{\op{S}}_{\mathrm{ext}}\{f\} (\mathbf r)
=
\junit k_{\mathrm{ext}}\eta_{\mathrm{ext}}
\int_S
\widetilde{G}_{\mathrm{ext}}(\mathbf r,\mathbf r')
f(\mathbf r')\,\mathrm{d}\mathbf r', && \mathbf r \in S\phantom{,}
\label{eq:Sout_m}
\end{flalign}
and
\begin{flalign}
\widetilde{\op{D}}_{\mathrm{ext}}^{*}\left\{f\right\} (\mathbf{r})
=
\int_S
\partial _n \widetilde{G}_{\mathrm{ext}}(\mathbf r,\mathbf r')
f(\mathbf r')\,\mathrm{d}\mathbf r', && \mathbf{r}\in S,
\label{eq:Dstar_gsie}
\end{flalign}
respectively, it is written as
\begin{equation}
\begin{bmatrix}
\widetilde{\op{S}}_{\mathrm{ext}}
&
\op{S}_{\mathrm{int}}
\\[4pt]
-\frac{1}{2}\op I
+
\widetilde{\op{D}}_{\mathrm{ext}}^{*}
&
\frac{1}{2}\op{I}
+
\op{D}_{\mathrm{int}}^{*}
\end{bmatrix}
\begin{bmatrix}
\widetilde{j}_{\mathrm{ext}}\\
\widetilde{j}_{\mathrm{int}}
\end{bmatrix}
=
\begin{bmatrix}
e_z^{\mathrm{inc}}\\
h_t^{\mathrm{inc}}
\end{bmatrix}.
\label{eq:gsie_system}
\end{equation}
In the corresponding discretized system, the system blocks $\widetilde{\mathbf S}_{\mathrm{ext}}$ and $\widetilde{\mathbf D}_{\mathrm{ext}}^{*}$ are associated only with the \ac{mgf} and are, therefore, asymptotically compressible, just like the moment matrices in \cite{dahan2024fast, kalhofer2025fast}. 
Their blocks $\widetilde{\mathbf S}_{\mathrm{ext}}^{\mathrm {\hat ss}}$ and $\widetilde{\mathbf D}_{\mathrm{ext}}^{\mathrm {\hat ss*}}$ are of ranks that grow very slowly (effectively $\mathcal O(1)$] with  $kR$. 
For lossy scatterers (complex-valued $k_{\mathrm{int}}$), the system blocks $\mathbf S_{\mathrm{int}}$ and $\mathbf D_{\mathrm{int}}^{*}$ are also compressible, thus making the entire system matrix in compressible. For lossless scatterers, the ranks for these blocks may still scale slowly with the electrical length until very large $kR$ values are reached, if $k_{\mathrm{ext}}\gg k_{\mathrm{int}}$. 

\section{Numerical Results}
\label{sec:numerical_results}
The following examples demonstrate the enhanced compressibility of the proposed formulation and validate it.
In all the examples, the materials are non-magnetic.
For discretization, pyramid bases on polygonal meshes and Galerkin testing are used. The coarsest discretization length used is $\delta_0=\lambda_{\min}/10$, where $\lambda_{\min}$ is the smallest between the wavelengths and skin-depth (for lossy scatterers) in $\Omega$ and $\bar \Omega$.
The \ac{gsie} uses multipole-based auxiliary contributions $G_{\mathrm{ext}}^{\mathrm{aux}}(\mathbf r, \mathbf r')$, as in~\cite{kalhofer2025fast}. For each $\mathbf r'$, let $\mathbf r''(\mathbf r')= \mathbf r'-\upsilon \hat {\mathbf {\zeta}}(\mathbf r')$ ($\hat {\mathbf {\zeta}}$ is the normal to the convex skull of $S$ at the point closest to $\mathbf r'$). From $\mathbf r''$, cylindrical contributions to $G_{\mathrm{ext}}^{\mathrm{aux}}(\mathbf r, \mathbf r')$ emanate. The stencil $G_{\mathrm{ext}}^{\mathrm{aux}}$ is defined in the local coordinate system centered at $\mathbf r''$, using $\rho_{\mathrm o}=|\mathbf r-\mathbf r''(\mathbf r')|$ and the angle $\varphi_{\mathrm o}$ between $\mathbf r$ and the axis connecting $\mathbf r'$ and $\mathbf r''(\mathbf r')$ (see Fig.~\ref{fig:greens_functions} inset), as 
\begin{equation}
G_{\mathrm{ext}}^{\mathrm{aux}}(\mathbf r, \mathbf r')=
\sum_{m=0}^{M-1}
a_m {\mathrm H}_m^{(2)}(k_{\mathrm{ext}}\rho_{\mathrm o})
\cos(m\varphi_{\mathrm o}).
\label{eq:multipole_aux}
\end{equation}
The multipole complex amplitudes $a_m$ are selected such that $\widetilde{G}_{\mathrm{ext}}(\mathbf r,\mathbf r')$ exhibits a deep shadow in $M'$ far field adjustment point in the range $\varphi_{\mathrm o}\in [\pi-\Delta_\varphi/2, \pi+\Delta_\varphi/2]$  (for complete detail, see \cite{kalhofer2025fast}). 
Additionally, $|G_{\mathrm{ext}}^{\mathrm{aux}}|$ is minimized near $\mathbf r'$ (see \cite{dahan2026reflective}). 
In the examples, the default settings are $\upsilon=0.3\lambda$, $\Delta_{\varphi}=120^\circ$, $M=13$, and $M'=11$. Whenever $\Delta_{\varphi}=150^\circ$ is used, $M=M'=14$. For lossless $\Omega$ and $\bar \Omega$, the wavenumber ratio is specified. The lossy $\Omega$ examples are of polyurethane [$\varepsilon_{\mathrm{int}}=3.7(1-5\cdot 10^{-2} \mathrm j)\varepsilon_0$] in air.
\begin{figure}[!b]
    \centering
    \includegraphics[width=\columnwidth]{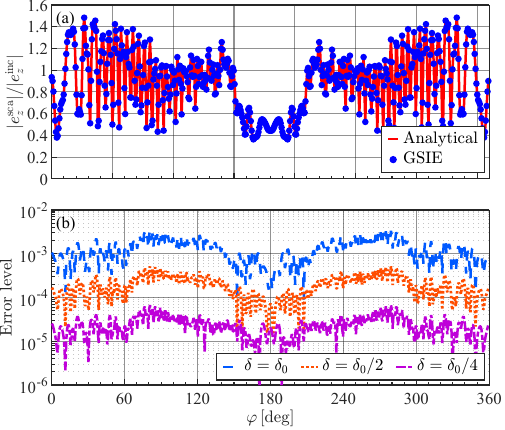}
    \caption{Circular dielectric scatterer. (a) Scattered field magnitudes. (b) Error level for different mesh sizes.}
    \label{fig:circle_results}
\end{figure}

In the first example, the rank-deficiency of moment matrix blocks for various kernels is demonstrated.  
In Fig.~\ref{fig:greens_functions}, the kernel magnitudes are presented for a source and observers on a circular contour of radius $R=40\lambda$. For $\Omega$, two cases are examined: a lossless material with $10k_{\mathrm{int}}=k_{\mathrm{ext}}$ and polyurethane in an air background. Two variants of $\widetilde{G}_{\mathrm{ext}}$ are examined. Both exhibit deep shadow compared to the conventional kernels in the lossless material domains, with a broader shadow region for $\Delta_{\varphi}=150^\circ$. Naturally, in the lossy medium, the kernel decays very fast as well. 
In Fig.~\ref{fig:compression_ranks}, for the kernels in Fig.~\ref{fig:greens_functions}, the scaling with $kR$ of the ranks (for a relative error threshold $\tau=10^{-4}$) of the blocks representing the interaction between a quarter of a circle of radius $R$ and its remainder is studied. For the lossless domains, the conventional kernel ranks scale fast. For the exterior domain, the rank approaches the expected $\mathcal O(kR)=\mathcal O(N)$ sooner. 
For conventional formulations, this behavior dominates the ranks of {\it all} system blocks, suggesting their limited asymptotic compressibility.
For the \ac{gsie} kernels, however, the rank remains roughly constant for broad $kR$-value ranges. This range broadens with $\Delta_{\varphi}$ (with a slight error increase that can be mitigated via the discretization parameters).
For the studied wavenumber ratio, for a broad range of  $kR$ values, the \ac{gsie} ranks are higher and govern the costs of solvers. The internal-domain ranks, which dominate beyond the crossover $kR$, are smaller than their external-domain counterparts by large factors.
For the lossy $\Omega$ case, the rank fluctuates before converging to a low constant value, allowing the compression to be dictated by the favorable scaling of the \ac{gsie} ranks.  

\begin{figure}[!b]
    \centering
    \includegraphics[width=\columnwidth]{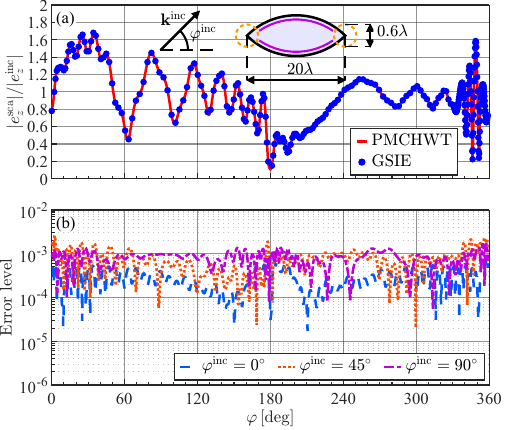}
    \caption{Dielectric ogive. (a) Scattered field magnitudes. (b) Error level for different incident angles  $\varphi^{\mathrm{inc}}$.}
    \label{fig:ogive_results}
\end{figure}

Next, the proposed formulation is validated via examination of the error in computation of the scattered fields, for unit $e_z^{\mathrm {inc}}$. This is done for the $ \Delta_{\varphi}=120^\circ$ configuration. The fields and errors are evaluated on surfaces placed $\lambda/4$ outside of $S$. 
The first case is of a lossless dielectric  circular cylinder with a radius $R=20\lambda$ and $k_{\mathrm{int}}=k_{\mathrm{ext}}/\sqrt{5}$. The \ac{gsie} results (see Fig.~\ref{fig:circle_results}) are compared to the analytical cylindrical function series solution~\cite{jin2015theory}. The error is controlled by the mesh size $\delta$. 
The second case is of an ogive with an end-to-end length of $20\lambda$ and the same material parameters. Here, for $\mathbf r'$ within a $\lambda/3$ distance of the ogive tips, where $\hat {\mathbf n}'$ is ill-defined, $G_{\mathrm{ext}}^{\mathrm{aux}}(\mathbf r,\mathbf r')$ was deactivated. A mesh length of $\delta_0/2$ was used. In Fig.~\ref{fig:ogive_results}, the \ac{gsie} solution is compared to that of the \ac{pmchwt} \ac{sie}, with the error evaluated for three incident angles. Similar error levels are observed for the three cases.

\begin{figure}[!t]
    \centering
    \includegraphics[width=\columnwidth]{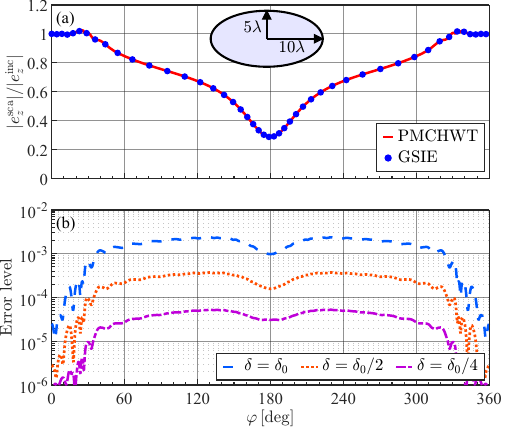}
    \caption{Lossy elliptical scatterer. (a) Scattered field magnitudes. (b) Error level for different mesh sizes.}
    \label{fig:ellipse_results}
\end{figure}

\begin{figure}[!b]
    \centering
    \includegraphics[width=\columnwidth]{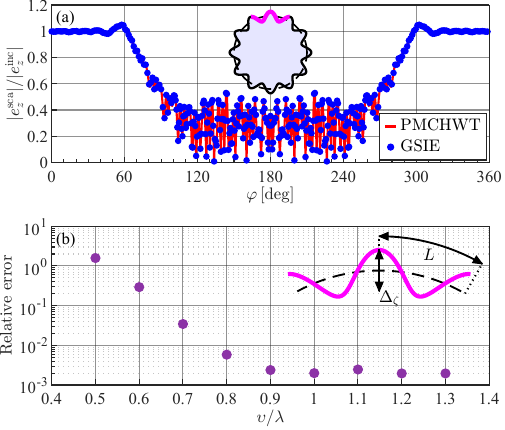}
    \caption{Lossy corrugated circular cylinder. (a) Scattered field magnitudes. (b) Relative 2-norm error as a function of  $\upsilon/\lambda$.}
    \label{fig:corrugated_results}
\end{figure}

In the last set of examples, lossy polyurethane scatterers are considered. The \ac{gsie} results are compared to those of the \ac{pmchwt} \ac{sie}. First, an ellipse with major and minor radii of $10\lambda$ and $5\lambda$ is examined. The error in Fig.~\ref{fig:ellipse_results} remains controllable with $\delta$. Finally, a corrugated circular scatterer, with a mean radius $R=10\lambda$ and corrugation height and cycle $\Delta_\zeta=\lambda$ and $L=1.96\lambda$, is studied. For $\delta = \delta_0/2$, the error is studied as a function of $\upsilon$ in Fig.~\ref{fig:corrugated_results}.
The 2-norm relative error is of levels similar to those of Fig.\ref{fig:ellipse_results} for $\upsilon>0.8\lambda $, suggesting that the formulation remains useful as long as $G_{\mathrm{ext}}^{\mathrm{aux}}(\mathbf r, \mathbf r')$ emanate from within (or close to) the convex skull of $S$.

\section{Conclusion}
 
The presented work extends the \ac{gsie} approach beyond impenetrable objects. The proposed formulation exhibits enhanced \ac{lr}-compressibility compared to the conventional formulations. This can be translated, by fast direct and iterative solvers, to great savings and, for certain classes of scatterers, asymptotically superior performance. The formulation shows similar levels of accuracy across ranges of scatterer material and geometric parameters. The multipole-based \ac{sie} formulation was shown to be easily adjustable to maintain the compressibility for electrically very large objects and to accurately treat geometries with isolated corners.
The formulation can be naturally integrated into multi-trace domain-decomposition solver architectures for more complex scatterers ~\cite{le2026multitracemuller, liang2025mixed}. Its \ac{te} and 3-D scalar (acoustic) and vector extensions will pave the way for designing penetrable object \acp{sie} that are both more compressible and better-conditioned~\cite{dahan2026wellconditioned}. The investigation of the well-posedness of \acp{gsie}, for both impenetrable and penetrable objects, is ongoing.

\section*{Acknowledgment}

The authors thank Yossi Dahan for his assistance in the GSIE kernel design.

\bibliographystyle{IEEEtran}
\bibliography{references}

\end{document}